# Chemical Potential and Quantum Hall Ferromagnetism in Bilayer Graphene

**Authors:** Kayoung Lee[1], Babak Fallahazad[1], Jiamin Xue[1], David C. Dillen[1], Kyounghwan Kim[1], Takashi Taniguchi[2], Kenji Watanabe[2], Emanuel Tutuc[1*]

**Affiliations:**

[1]Microelectronics Research Center, The University of Texas at Austin, 10100 Burnet Rd., Austin, TX 78758, USA.

[2]National Institute for Materials Science, 1-1 Namiki Tsukuba Ibaraki 305-0044, Japan.

*Correspondence to: etutuc@mer.utexas.edu

**Abstract**:

Bilayer graphene has a unique electronic structure influenced by a complex interplay between various degrees of freedom. We probe its chemical potential using double bilayer graphene heterostructures, separated by a hexagonal boron nitride dielectric. The chemical potential has a non-linear carrier density dependence, and bears signatures of electron-electron interactions. The data allow a direct measurement of the electric field-induced bandgap at zero magnetic field, the orbital Landau level (LLs) energies, and the broken symmetry quantum Hall state gaps at high magnetic fields. We observe spin-to-valley polarized transitions for all half-filled LLs, as well as emerging phases at filling factors $\nu = 0$ and $\nu = \pm 2$. Furthermore, the data reveal interaction-driven negative compressibility and electron-hole asymmetry in $N = 0, 1$ LLs.

**Main Text:**

The electronic structure of bilayer graphene has a bandgap and energy-momentum dispersion that can be tuned by an applied transverse electric field (*1-4*). In high magnetic fields, electron-electron interaction coupled with the spin and valley degrees of freedom can lead to a diverse set of quantum Hall states (QHSs) (*1-8*). Thermodynamic measurements of the chemical potential or the density of states are fundamental to understanding the bilayer graphene electronic properties (*9-12*). The density of states (DOS) of bilayer graphene extracted from compressibility data (*9-10*) shows modulations associated with four-fold, spin and valley degenerate Landau levels (LLs), whereas local compressibility measurements in single-gated suspended bilayer graphene samples reveal broken symmetry QHSs (*11*).

Here we study samples consisting of two Bernal stacked bilayer graphene, separated by a thin hexagonal boron nitride (hBN) (Fig. 1A). The samples are fabricated using a sequence of mechanical exfoliation, dry-transfer, and alignment with a pre-existing device, similarly to refs *13*, *14*. The bottom bilayer graphene is supported by a thick (30-50 nm) hBN substrate, mechanically exfoliated on a $SiO_2$/Si substrate. A relatively thin (2-6 nm) hBN layer is then transferred and aligned with the bottom bilayer, followed by the top bilayer graphene transfer. Each bilayer graphene is patterned into a multi-terminal Hall bar, and an etch-through of the two bilayers ensures the Hall bar edges are aligned (Fig. 1B, inset). The bilayer resistances are measured as a function of back-gate ($V_{BG}$), and inter-bilayer bias applied to the top bilayer ($V_{TL}$) using small signal, low frequency lock-in techniques. Three samples, labelled #1, #2, and #3, were investigated in this study. The bottom bilayers in all the samples have mobilities ranging between 100,000 – 290,000 $cm^2$/Vs, whereas the top bilayer graphene have lower mobilities, ranging between 3,400 – 6,500 $cm^2$/Vs.

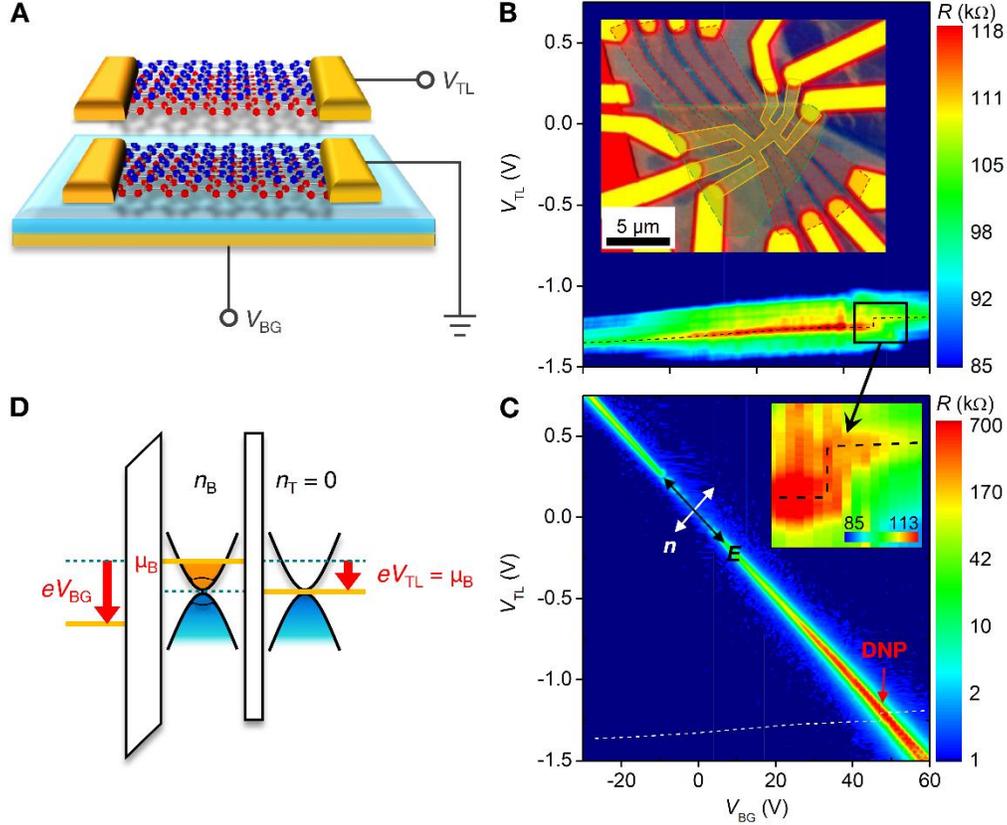

**Fig. 1**. **Sample schematic and characterization**. (**A**) Schematic of the double bilayer heterostructure. Top (**B**), and bottom (**C**) bilayer resistance ($R$) measured as a function of $V_{TL}$ and $V_{BG}$. Dashed lines in the main plots of (B) and (C) mark the charge neutrality of the top bilayer. The bottom bilayer $n$ and $E$-field axes are marked in (C). Inset in (B) shows a false color optical micrograph of the device. The yellow (dashed red) contour marks the top (bottom) bilayer; the dashed green line marks the inter-bilayer hBN perimeter. Inset in (C) shows a magnified view of the top bilayer charge neutrality line near the double neutrality point. (**D**) Energy band profile across the heterostructure with the top layer at charge neutrality.

The top ($n_T$) and bottom ($n_B$) bilayer carrier densities depend on $V_{BG}$ and $V_{TL}$ as (*15*):

$$eV_{BG} = e^2(n_B + n_T)/C_{BG} + \mu_B \quad (1)$$

$$eV_{TL} = -e^2 n_T / C_{int} - \mu_T + \mu_B \quad (2)$$

Here, $C_{BG}$ and $C_{int}$ represent the bottom and inter-bilayer dielectric capacitances, whereas $\mu_T$ and $\mu_B$ represent the chemical potentials (Fermi energies) of the top and bottom bilayers, respectively; $e$ is the electron charge. We note that $\mu$ and $n$ are positive (negative) for electrons

(holes), and $V_{BG}$ and $V_{TL}$ in Eqs. 1 and 2 are referenced with the bias values at which both bilayers are charge neutral, i.e. at the double neutrality point (DNP). The externally applied transverse electric field ($E$) across the bottom bilayer can be written as:

$$E = en_B/2\varepsilon_0 + en_T/\varepsilon_0 + E_0, \quad (3)$$

where $\varepsilon_0$ is the vacuum permittivity, $n_B$ and $n_T$ are determined using Eqs. 1 and 2, and $E_0$ is an additive constant which allows for a finite $E$-field at the DNP, associated with an unintentional doping of the top bilayer.

The measured bottom bilayer density and resistance dependence on $V_{BG}$ and $V_{TL}$ (Fig. 1C) is similar to that of a dual gated bilayer (*16-18*), where the charge neutrality point has a linear dependence on $V_{BG}$ and $V_{TL}$, with a slope controlled by $C_{BG}$ and $C_{int}$. Along the charge neutrality line of the bottom bilayer the resistance is minimum at $E = 0$, and increases with the $E$-field thanks to the $E$-field-induced bandgap of bilayer graphene. By comparison to the bottom bilayer, the top bilayer graphene resistance (Fig. 1B) has a weak dependence on $V_{BG}$ because of the screening by the bottom bilayer, and is controlled primarily by $V_{TL}$. Interestingly, setting $n_T = 0$ in Eq. 2 yields $eV_{TL} = \mu_B$, which implies that the inter-bilayer bias required to bring the top bilayer to charge neutrality, marked by dashed lines in Figs. 1B and 1C, is simply the chemical potential of the bottom bilayer (Fig. 1D) in units of eV (*15*). Using Eq. 1 the $n_B$ values along the top bilayer neutrality line (dashed lines in Figs. 1B and 1C) are $n_B = C_{BG} \cdot (V_{BG} - V_{TL})/e$. Consequently, the bottom bilayer chemical potential can be mapped as a function of its carrier density.

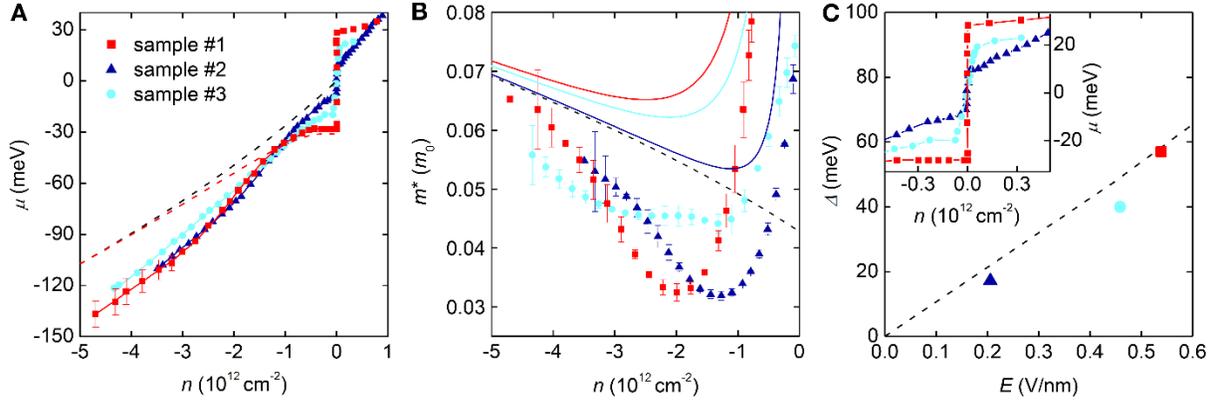

**Fig. 2. Chemical potential and effective mass dependence on carrier density.** (**A**) Bottom bilayer $\mu$ vs. $n$ in samples #1-3. Dashed lines: $\mu$ vs. $n$ calculated using the tight binding model for $E = 0$ (black) and $E = 0.54$ V/nm (red). (**B**) Symbols: Measured $m^*$ in units of bare electron mass ($m_0$) vs. $n$. Also displayed are the $m^*$ vs. $n$ calculated at $E = 0$ (black dashed), 0.54 V/nm (red), 0.21 V/nm (dark blue) and 0.46 V/nm (light blue). (**C**) Measured (symbols) and calculated (*19*) (dashed line) $\Delta$ values vs. $E$-field. Inset shows a magnified view of the $\mu$ vs. $n$ near charge neutrality.

Figure 2A shows the bottom bilayer chemical potential ($\mu$) vs. density ($n$) determined as described above. The finite doping of the top bilayer at $V_{TL} = V_{BG} = 0$ V in our samples leads to a finite $E$-field across the bottom bilayer at the DNP: $E_0 = 0.54$ V/nm, 0.21 V/nm, and 0.46 V/nm for samples #1, #2, and #3, respectively. From Eqs. 1-3 it follows that $E_0 = C_{BG} \cdot \Delta V_{BG} / \varepsilon_0$, where $\Delta V_{BG}$ is the difference between the back-gate bias at the DNP and at $n_B = 0$ and $E = 0$ point (Fig. 1C). The $\mu$ vs. $n$ for the lowest energy band in bilayer graphene calculated within a tight binding approximation (*3*) for $E = 0$ and $E = 0.54$ V/nm are included for comparison (dashed lines), using a non-interacting in-plane velocity $\upsilon = 8.4 \times 10^5$ m/s, inter-layer hopping $\gamma_1 = 0.34$ eV (*19*), and neglecting trigonal warping. Because of the interaction-induced renormalization of electron energies (*20*), the measured $|\mu|$ values are, particularly at high densities, larger than the results of the band calculations. The nonlinear $\mu$ vs. $n$ dependence indicates a strongly non-parabolic energy-momentum dispersion (*3, 4*). Such a dispersion would lead to a density-dependent effective mass, as observed in ref. *21*. This is confirmed in Fig. 2B, which shows a non-monotonic dependence of the effective mass $m^*$ on density $n$, extracted from Fig. 2A data using $m^* = (\pi\hbar^2/2)(d\mu/dn)^{-1}$. At low densities $m^*$ increases with the $E$-field, and shows a divergence as a function of $n$ near charge neutrality. The measured $\mu$ value also shows a clear discontinuity as $n$ changes sign, revealing a gap $\Delta$ (Fig. 2C) associated with the $E$-field-induced band-gap in bilayer graphene (*3, 4*).

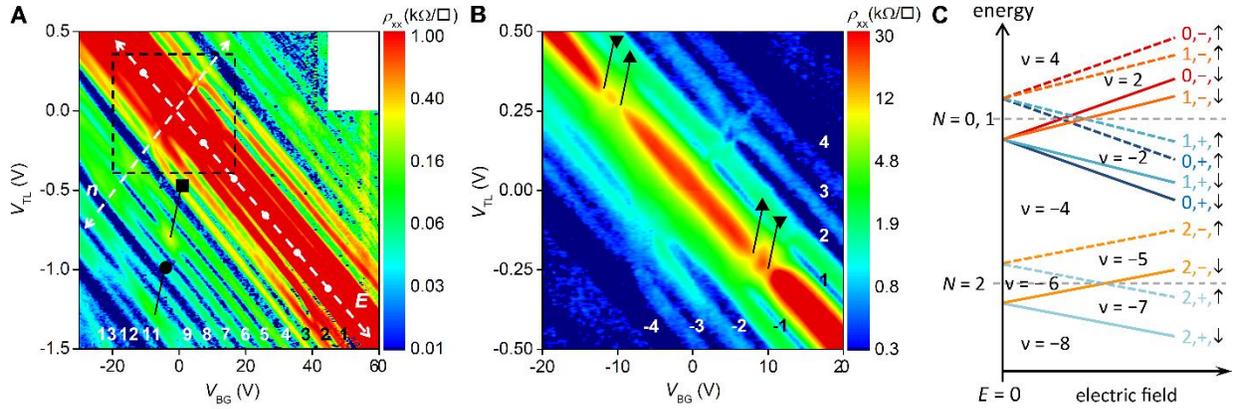

**Fig. 3**. **Bilayer graphene quantum Hall ferromagnetism.** (**A**) The bottom bilayer resistivity $\rho_{xx}$ as a function of $V_{TL}$ and $V_{BG}$, at $B = 14$ T and $T = 1.5$ K in sample #1. The dashed white lines mark the $n = 0$ and $E = 0$ axes; the dots along the $n = 0$ axis mark $E$-field increments of 0.1 V/nm. The absolute values of the QHS filling factors are shown. The $E$-field controlled spin-to-valley polarization transitions in the $N = 2$ (square), and $N = 3$ (circle) LLs are marked. (**B**) Magnified view of the dashed line rectangle of panel (A). The triangles mark two distinct transitions of the $\nu = 0$ QHS as a function of $E$-field. (**C**) Schematic representation of the LL evolution with $E$-field, and the ensuing QHSs. The solid (dashed) line marks the spin down (up) levels. The orbital index, and layer (+,−) degrees of freedoms are color coded. Assuming the direction of the applied $E$-field favors energetically the occupation of the bottom layer, the symbols +, − mark the bottom and top layers, respectively.

In a perpendicular magnetic field ($B = 14$ T), sample #1 shows clear minima of the longitudinal resistivity $\rho_{xx}$ (Fig. 3A), indicating QHSs at all integer filling factors ($\nu$) up to $|\nu| = 15$; the $\nu = 0$ QHS is marked by a $\rho_{xx}$ maximum. The QHSs at $\nu \neq \pm 4, 8, 12, \ldots$ stabilized by the spin and valley degeneracy lifting induced by the interaction-enhanced Zeeman splitting (*6*) and the *E*-field, respectively, undergo transitions at finite *E*-field values as a result of the interplay between the LL spin and valley splitting. Specifically, the $\nu$ = -1, -3, -5, -7, -9, -11 (odd filling) QHSs are absent at $E = 0$ and emerge as an *E*-field is applied. On the other hand the half fillings $\nu$ = 0, -6, -10 of the spin and valley degenerate LLs are present at $E = 0$, and collapse at finite *E*-field values. The interplay between the spin and valley splitting explains the transitions at $\nu$ = -6, -10, -14 at a finite *E*-field (Fig. 3A and supplementary text).

Figures 3B and S3 show a strong $\nu = 0$ QHS at $E = 0$, as well as at high *E*-field, consistent with refs. *17, 18*. Both theory and experiment (*7, 22*) indicate that the insulating $\nu = 0$ QHS in the proximity of $E = 0$ is described by a canted antiferromagnetic (CAF) ground state, where electrons in different valleys have opposite in-plane spin orientation and a net out-of-plane spin polarization; the $\nu = 0$ state at high *E*-field is valley (layer) polarized. In Fig. 3B, the insulating $\rho_{xx}$ at $\nu = 0$ collapses at two distinct *E*-field values, rather than one (*17, 18*), indicating two distinct transitions and the observation of an intermediate phase in between the CAF and layer-polarized phases. The data can be qualitatively understood using the tight-binding LL energy diagram (Fig. 3C, Ref. *3*). In this picture the energies of LL with orbital index $N = 0, 1$ have a different dependence on the *E*-field, which explains the two distinct transitions at $\nu = 0$, as well as the emergence of $\nu = \pm 1$ and $\pm 3$ QHSs at a finite *E*-field. When electron-electron interactions are included (*8*), the intermediate $\nu = 0$ QHS between the CAF and layer-polarized phases is found to be a more subtle spin-layer coherent phase, where LLs with the same $N$ but different spin and valley degrees of freedom, e.g. solid red (orange) and dashed dark (light) blue in Fig. 3C form coherent superpositions. In the single-particle picture, at $\nu = 0$ electrons in the $N = 0$, spin down, top layer LL (solid red line in Fig. 3C) should move to the $N = 0$, spin up, bottom layer LL (dashed dark blue) at a finite $E$-field by changing both spin and valley orientations, while retaining the orbital index. In the many-body picture however, electrons favor a coherent superposition of the two states at the transition between the CAF and layer-polarized phases. The $\nu = \pm 2$ QHSs are present at $E = 0$, vanish at a small $E$-field, and then reemerge (Figs. 3B and S3). The $\nu = \pm 2$ at $E = 0$ can be explained as layer-coherent QHSs, where the LLs with the same $N$ and spin orientation but different layer (valley) degrees of freedom form coherent superpositions (*8*).

Through the chemical potential mapping we determine the LL energies in bilayer graphene as a function of $\nu$ and $B$-field. Figure 4A shows the bottom bilayer $\rho_{xx}$ measured at $B = 12$ T, with the top bilayer charge neutrality line transposed on the contour plot (white line). Note that the transverse $E$-field across the bottom bilayer at DNP is 0.54 and 0.21 V/nm for sample #1 and #2 respectively, which lead to that the bottom bilayer $\nu = 0$ QHS at the DNP is layer polarized (Supplementary text and Fig. S4). At each integer filling of the bottom bilayer the top bilayer neutrality line displays an abrupt change in the $V_{TL}$ value, which translates into a chemical potential jump of bottom bilayer. The staircase-like dependence, particularly sharp at $\nu = 0, -4, -8, -12, -16$, testifies to a reduced LL broadening, in contrast to previous measurement in double monolayer heterostructures using metal-oxide as interlayer dielectric (*15*).

Figure 4B represents the LL orbital energies as a function of $B$-field and $N$, determined from the chemical potential at the half filling of each LL orbital index. The LL orbital energies increase linearly with $B$-field, consistent with the theoretical $\mu = \hbar \omega_c \sqrt{N(N-1)}$ dependence (*3*); here $\omega_c = eB/m^*$ is the cyclotron frequency. The inset of Fig. 4B shows the $m^*$ vs. $N$ dependence determined from the slope of $\mu$ vs. $B$ at each LL. The increasing $m^*$ with $N$ is similar to $B = 0$ T data of Fig. 2B.

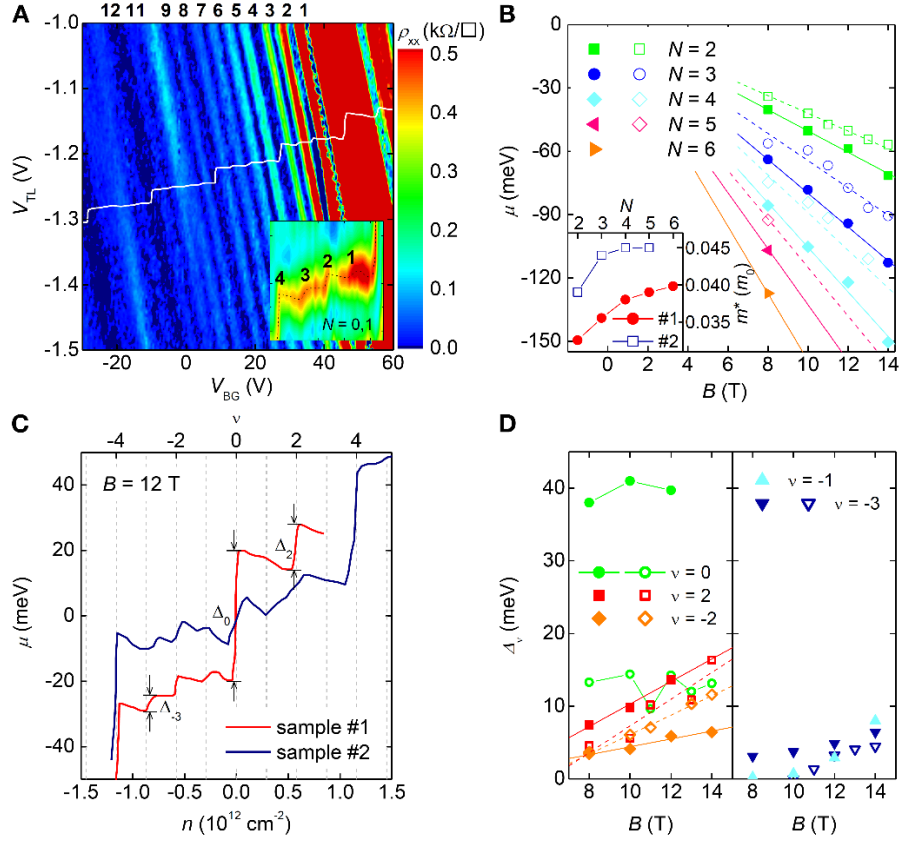

**Fig. 4. Landau level energies and broken symmetry gaps in bilayer graphene.** (**A**) The bottom bilayer resistivity $\rho_{xx}$ as a function of $V_{TL}$ and $V_{BG}$ at $B = 12$ T and $T = 1.5$ K in sample #1, along with the top bilayer charge neutrality locus (white line). The top axis shows the absolute values of the QHS filling factors. The scale bar corresponds to main panel data. Inset shows the top bilayer resistance dependence on $V_{TL}$ and $V_{BG}$, as the top bilayer charge neutrality line crosses the $\nu = -4, -3, -2, -1, 0$ QHSs of the bottom bilayer; data ranges from 270 k$\Omega$ (blue) to 350 k$\Omega$ (red). (**B**) Orbital LL energies as a function of $B$-field and LL index in samples #1 (filled symbols) and #2 (open symbols). The inset shows the $m^*$ vs. $N$ extracted from the main panel data. (**C**) Bottom bilayer $\mu$ vs. $n$ (bottom axis) and $\nu$ (top axis) at $B = 12$ T. (**D**) QHS gaps at $\nu = -3, -2, -1, 0, 2$ in sample #1 (filled symbols) and #2 (open symbols).

The chemical potential of $N = 0, 1$ LLs, shown in Fig. 4A inset and Fig. 4C, reveals jumps at integer fillings $\nu = -3, -2, -1, 0$, and 2, as well as a decreasing $\mu$ vs. $\nu$ (or $n$) in the proximity of the QHSs stabilized in the $N = 0, 1$ LLs. The decreasing $\mu$ vs. $n$ dependence stems from a strong exchange interaction, and translates into a negative compressibility of the electron system (*23*). A similar observation has been reported in monolayer graphene (*24*).

Using the chemical potential jump at integer fillings, we determine the broken symmetry QHS gaps ($\Delta_\nu$) at filling factors $\nu = -3, -2, -1, 0, 2$, as a function of $B$-field (Fig. 4D). We note that the measured $\Delta_0$ is independent of $B$-field, in contrast to the linear $B$-field dependence of $\Delta_0$ observed in single-gated bilayer graphene (*11, 25*). However, Fig. 4D data represent $\Delta_0$ in the layer polarized phase, whereas refs. *11* and *25* probe $\Delta_0$ in the CAF phase. The layer polarized $\Delta_0$ is controlled by the interaction and Zeeman splitting, as well as the $E$-field induced on-site energy difference. At moderate $E$-fields the competition between the single particle splitting, which decreases with the $B$-field, and the $\propto \sqrt{B}$ interaction term leads to a $\Delta_0$ weakly dependent on $B$-field, and mainly controlled by the $E$-field (*8*). This explains the larger (smaller) $\Delta_0$ value in sample #1 (#2), thanks to the larger (smaller) bottom layer $E$-field near the DNP.

The gaps of $\nu = -3, -1, -2$, and 2 have a linear dependence on the $B$-field, a trend similar to results obtained in single-gated suspended bilayer graphene (*11*). Whereas theoretical considerations (*6, 8*) suggest a sublinear $B$-field dependence associated with the $\propto \sqrt{B}$ of the interaction energy, the non-linearity is weak particularly in the $B$-field range probed here. Furthermore, the broken symmetry QHSs observed in the $N = 0, 1$ LLs show a marked electron-hole asymmetry. Specifically, $\Delta_2$ is larger than $\Delta_{-2}$ for both samples #1 and #2, whereas $\Delta_{-1}$ and $\Delta_{-3}$ are larger than $\Delta_1$ and $\Delta_3$, respectively; $\Delta_1$ and $\Delta_3$ are too small to be resolved experimentally. The electron-hole asymmetry and the differences in $\Delta_2$ and $\Delta_{-2}$ in the two samples with different applied $E$-fields (Fig. 4D) observed experimentally are qualitatively consistent with a detailed Hartree-Fock analysis (*8*) of the broken symmetry QHSs of the $N = 0, 1$ LLs. .

**Acknowledgments:** This work was supported by ONR, NRI SWAN, and Intel. We thank A. H. MacDonald, R. Côté, P. Kim, and F. Ghahari for discussions.

**Supplementary Materials**
Materials and Methods
Supplementary Text
Figs. S1 to S4
Table S1
Full reference list

# Supplementary Materials for

## Chemical Potential and Quantum Hall Ferromagnetism in Bilayer Graphene

Kayoung Lee, Babak Fallahazad, Jiamin Xue, David C. Dillen, Kyounghwan Kim,
Takashi Taniguchi, Kenji Watanabe, Emanuel Tutuc

**Materials and Methods**

The fabrication of samples investigated here starts with consecutive mechanical exfoliations of hBN onto silicon dioxide thermally grown on a highly doped Si substrate, and bilayer graphene flakes onto poly (methyl methacrylate) layer on a water-soluble polyvinyl alcohol. AFM was performed on hBN to determine the thickness, and probe the surface topography, and optical contrast and Raman spectroscopy were used to identify the number of graphene layers. The bilayer graphene was transferred on top of the hBN using the dry-transfer and alignment methods similar to those described in refs. (*13*) and (*14*). Electron-beam lithography and oxygen plasma etching were performed to define a multi-terminal Hall bar. The second, relatively thin hBN crystal and the topmost bilayer graphene ware successively transferred on top of the Hall bar, and the top graphene was patterned into a multi-terminal Hall bar. An etch-through of the top and bottom graphene layers ensures the Hall bar edges are aligned. After the transfer of each layer, ultra high vacuum annealing at either 350 °C or 320 °C was performed to remove process residue.

Two and four-point resistance measurements were performed on the top and bottom bilayer graphene by flowing source currents of 1 nA on each layer, using lock-in amplifiers as described in Fig. S1. Different lock-in frequencies, ranging between 11 - 17 Hz were chosen for the top and bottom bilayers to exclude cross-talk. A radio-frequency transformer (Jensen Transformers, model JT-SUB-BB) is used to flow an AC current on the top layer, while applying a DC bias $V_{TL}$, with respect to ground. The samples were measured in a variable-temperature liquid $^4$He flow cryostat in zero and high magnetic fields up to 14 T.

**Supplementary Text**

Comparison of the bilayer graphene chemical potential with tight-binding calculations

The electronic properties of bilayer graphene can be conveniently captured by the Slonczeweski-Weiss-McClure model (*26*). In this model, the most important parameters are the nearest-neighbor hopping amplitude, which determines the Fermi velocity in mono-layer graphene, and the nearest-neighbor interlayer hoping amplitude $\gamma_1$. Electron-electron interaction can reshape the energy-momentum dispersion for both monolayer (*27*) and bilayer graphene (*20*). In monolayer graphene the interaction leads to a density dependent Fermi velocity (*24, 28, 29*). The most recent experimental study (*24*) indicates a non-interacting Fermi velocity $v_{F0} = 8.5 \times 10^5$ m/s (*19, 24, 28*) in monolayer graphene. The values for $\gamma_1$ range between 0.3 eV (*30*) and 0.4 eV (*31, 32*), with the most recent experimental study suggesting $\gamma_1 = 0.38$ eV (*33*). The latter value is close to $\gamma_1 = 0.36$ eV reported by a most recent *ab initio* tight binding model (*34*). To compare the measured $\mu$ vs. $n$ data in bilayer graphene with tight-binding calculations (Fig. S2), we use $v_{F0} = 8.4 \times 10^5$ m/s (*19, 24, 28*), and $\gamma_1$ values of 0.34 eV (*19*) and 0.4 eV (*31, 32*). In both cases the measured $\mu$ vs. $n$ data are larger than tight-binding calculations, an observation attributed to interaction-induced renormalization of the bilayer graphene energy-momentum dispersion (*20*).

Quantum Coherent phases at $\nu = \pm 2$ QHSs

The $\nu = \pm 2$ QHSs are present at $E = 0$, vanish at a relatively small $E$-field, and then reemerge (Fig. 3B and Fig. S3). The $\nu = \pm 2$ at $E = 0$ can be explained as a layer-coherent QHS, where the LLs with the same orbital index and spin orientation but different layer (valley) degrees of freedom, e.g. dashed red (orange) and dashed dark (light) blue LLs of Fig. 3C form a coherent superposition thanks to the exchange interaction (*8*). At a finite $E$-field the different on-site energies in different layers lead to incoherent $\nu = \pm 2$ phases with a zero in-plane and a net out-of-plane spin polarizations.

Spin-to-valley polarized transitions

The interplay between the spin splitting and the $E$-field induced valley (layer) splitting at $N = 2$ depicted in Fig. 3C explains the absence of $\nu = -5$ and $-7$ QHSs at $E = 0$, their emergence with the applied $E$-field, as well as the spin-to-valley polarized phase transition at $\nu = -6$ at a finite $E$-field (Fig. 3A). Similar phase transitions are observed at $\nu = -10$ and $-14$, half-filled LL QHSs.

Figure S4 summarizes the critical $E$-field values at which half filled LL QHSs undergo the spin-to-valley polarized phase transitions discussed above. The $E$-field values are calculated using Eq. 3, and including the layer Fermi energies. Because the $N = 0, 1$ LL wave-functions in different valleys are fully layer polarized (*3*) we use for comparison the internal electric field in a layer polarized $\nu = 0$ QHS, $E_{int} = 4(e^2B/h)/2\varepsilon_0$ (dashed line) (*18*). The critical $E$-field values at which the LLs become layer polarized are comparable to $E_{int}$ for $N = 0$ and $N = 1$, and increase for higher $N$ as higher LL wave-functions have a reduced layer polarization by comparison to $N = 0, 1$ (*5*).

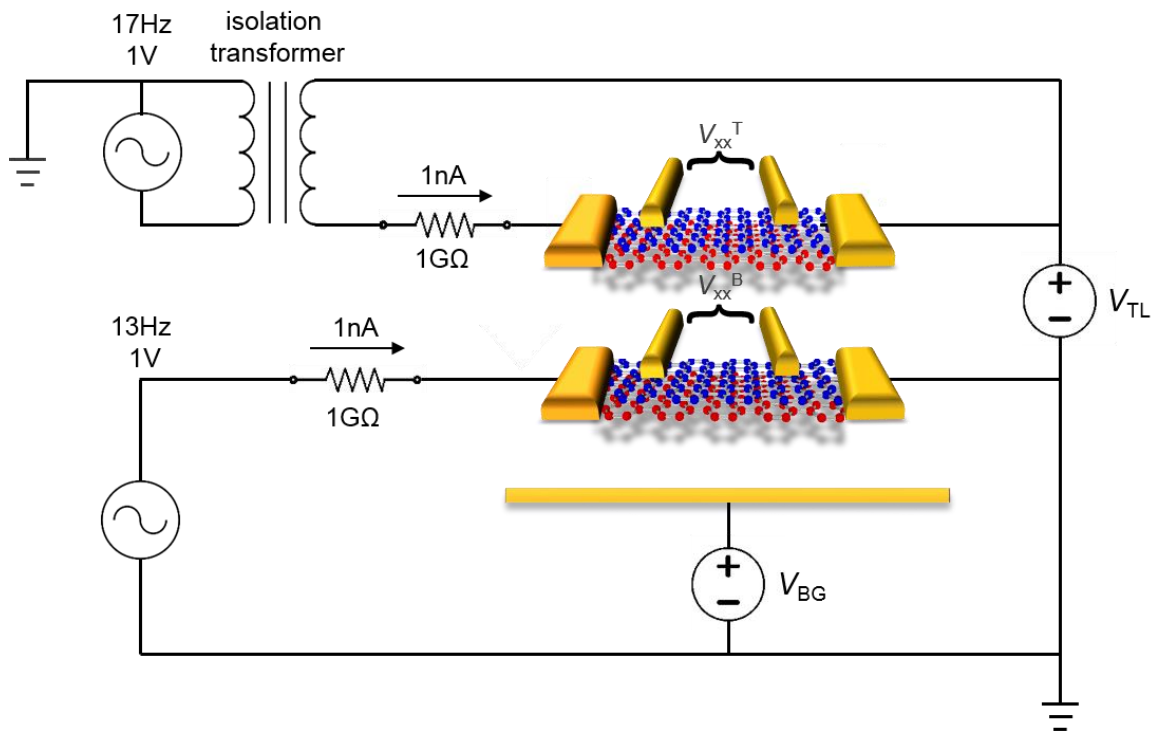

**Fig. S1**
Schematic of the circuit used to probe the double bilayer as a function of interlayer ($V_{TL}$) and back-gate ($V_{BG}$) bias. The two bilayers are separated for clarity.

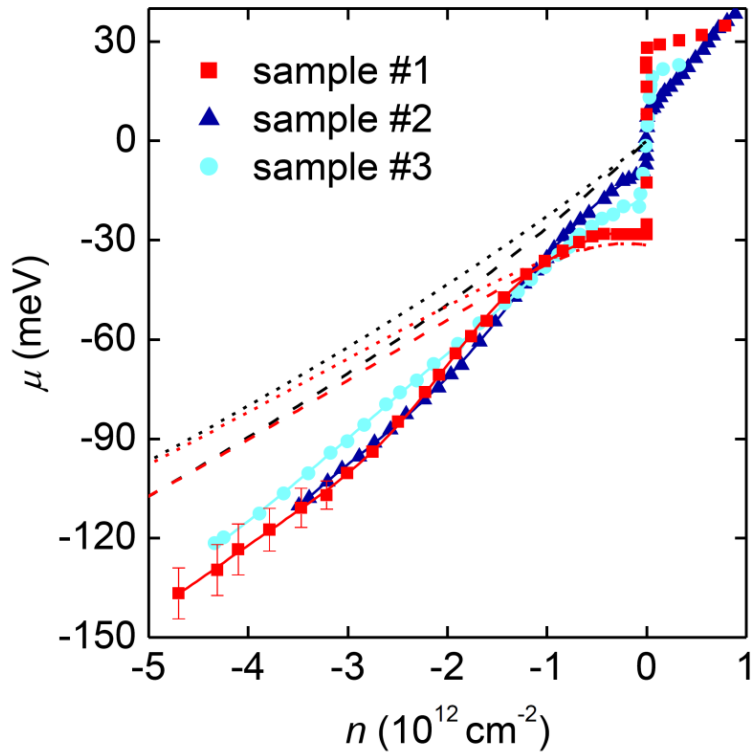

**Fig. S2**

Bottom bilayer $\mu$ vs. $n$ measured in samples #1-3 (symbols) compared with tight-binding calculations (lines) using at $E = 0$ (black) and $E = 0.54$ V/nm (red). We used a non-interacting in-plane velocity of $8.4 \times 10^5$ m/s (*19, 24, 28*), and two values for the inter-layer hopping $\gamma_1 = 0.34$ eV (*19*) (dashed) or 0.4 eV (*29*) (dotted).

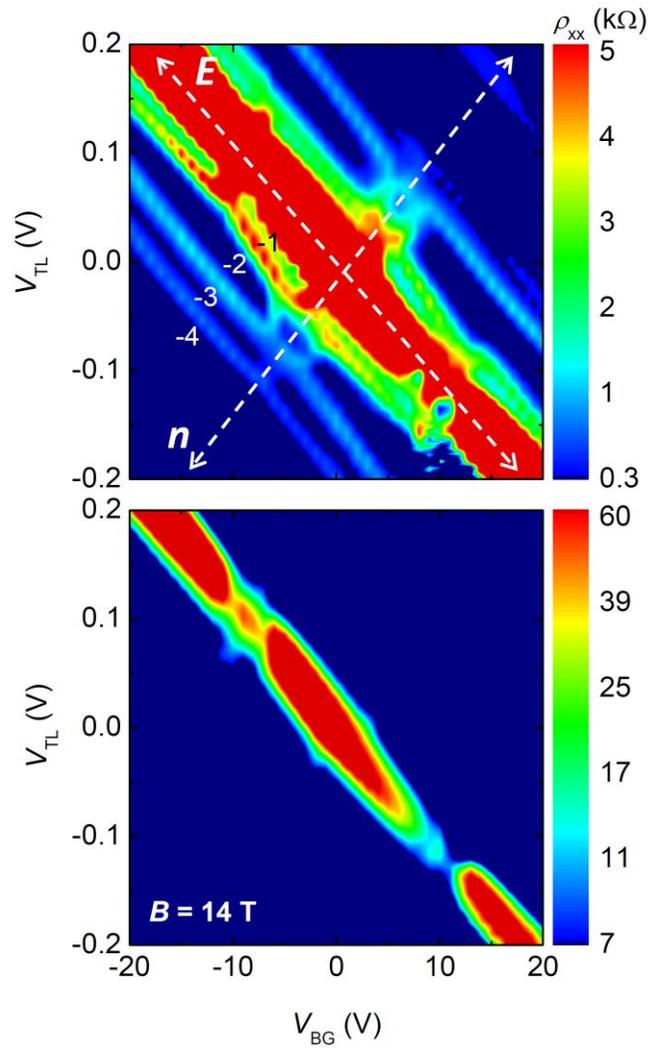

**Fig. S3**

Contour plot of the bottom layer resistivity $\rho_{xx}$ as a function of $V_{TL}$ and $V_{BG}$, measured at $B = 14$ T and $T = 1.4$ K in sample #2. Upper and lower panels show the same data, but in different ranges to better illustrate the evolution of $\nu = -1, -2, -3$ QHSs (upper panel), and the insulating $\nu = 0$ QHS (lower panel) as a function of $E$-field. QHSs at integer filling factor $\nu = -1, -2, -3$ are marked by local $\rho_{xx}$ minima (upper panel), and the QHS at $\nu = 0$ is marked by a $\rho_{xx}$ maximum (lower panel). The dashed white lines in the upper plot represent the $n = 0$, and $E = 0$ axes.

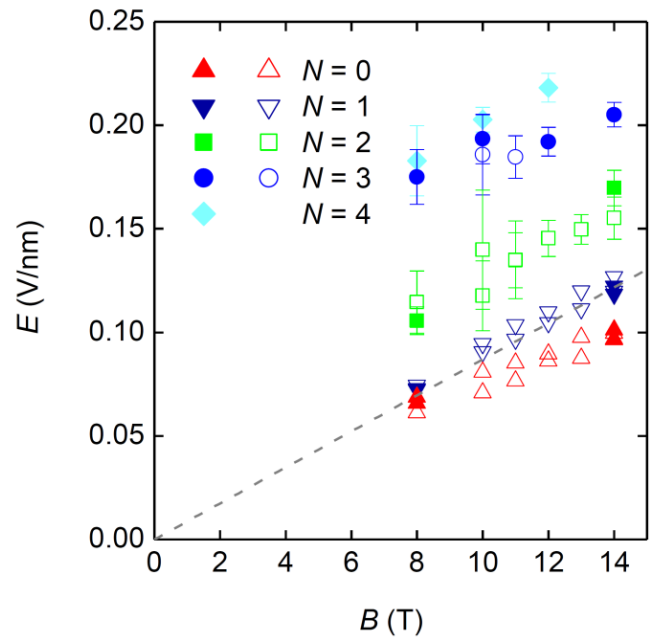

**Fig. S4**

$E$-field vs. $B$ at which QHS residing in different orbital LLs undergo spin-to-valley polarized transitions in sample #1 (filled symbols), and #2 (open symbols). The $N = 0, 1$ transitions are defined using Fig. 3C LL diagram.

**Table S1.**

Thickness of interlayer hBN, and bottom ($C_{BG}$) and inter-bilayer ($C_{int}$) dielectric capacitances for samples investigated here.

|  | Thickness of interlayer hBN | $C_{BG}$ (nF/cm$^2$) | $C_{int}$ (nF/cm$^2$) |
|---|---|---|---|
| Sample #1 | 5 – 5.5 nm | 9.8 | 384 |
| Sample #2 | 2 nm | 9.55 | 860 |
| Sample #3 | 5 nm | 10.7 | 410 |